# Group velocity locked vector dissipative solitons in a high repetition rate fiber laser

Yiyang Luo[1], Deming Liu[1,*], Lei Li[2], Qizhen Sun[1,*], Zhichao Wu[1], Zhilin Xu[1], Songnian Fu[1], and Luming Zhao[2]

[1]School of Optical and Electronic Information, National Engineering Laboratory for Next Generation Internet Access System, Huazhong University of Science and Technology, Wuhan 430074, Hubei, P. R. China;
[2]Jiangsu Key Laboratory of Advanced Laser Materials and Devices, School of Physics and Electronic Engineering, Jiangsu Normal University, Xuzhou, 221116 Jiangsu, P. R. China.
[*]Corresponding author: dmliu@mail.hust.edu.cn; qzsun@mail.hust.edu.cn.

**Abstract:** Vectorial nature of dissipative solitons (DSs) with high repetition rates is studied for the first time in a normal-dispersion fiber laser. Despite the fact that the formed DSs are strongly chirped and the repetition rate is greater than 100 MHz, polarization locked and polarization rotating group velocity locked vector DSs can be formed under 129.3 MHz fundamental mode-locking and 258.6 MHz harmonic mode-locking of the fiber laser, respectively. The two orthogonally polarized components of these vector DSs possess distinctly different central wavelengths and travel together at the same group velocity in the laser cavity, resulting in a gradual spectral edge and small steps on the optical spectra, which can be considered as an auxiliary indicator of the group velocity locked vector DSs.

**Key words:** Pulse propagation and temporal solitons; Nonlinear optics, fibers; Mode-locked lasers; Lasers, fiber.

## 1. Introduction

Dissipative solitons (DSs) in passively mode-locked fiber lasers have been a topic of intensive research as a fundamental extension of conservative solitons [1-3]. Different from the simple balance between nonlinearity and dispersion in conservative systems, continuous flow of energy in dissipative systems, namely a composite balance between the cavity dispersion, fiber nonlinear Kerr effect, laser gain and loss of fiber lasers plays a significant role in the formation of DSs. On one hand, DSs as an attractor are actually released from the constraint of initial conditions and fixed with a given set of laser parameters, thereby guaranteeing their practical applications for pulse generation, all-optical sampling and optical communications [1]. On the other hand through tuning the laser parameters, several fascinating behaviors such as vector dissipative solitons (VDSs) [4-7], dispersion-managed DSs [8-10], DS molecules [11-14], and dissipative soliton resonance (DSR) [15, 16] have been witnessed with varieties of laser configurations. Thus it can be seen that passively mode-locked fiber lasers with innovative laser designs serve as fertile playground for exploring the dynamics of DSs and extending their potential applications.

Recently, vectorial nature of DSs is taken into account for extending the exploration of the dynamics of DSs, as well as its promising applications of expanding the optical communication capacity based on polarization division multiplexing (PMD) and polarization switch (PS) [17]. VDSs composed of two orthogonally polarized DSs have been theoretically predicted and experimentally demonstrated when no polarization sensitive components are utilized in the laser cavity. Obviously, they are characterized by more complex behaviors and much richer dynamics than their scalar, one-component counterparts. Traditional research of vectorial nature of DSs is focused on conventional Schrödinger solitons, namely DSs formed in anomalous-dispersion regime, principally including group velocity locked vector Schrödinger solitons [4, 18, 19], polarization locked vector Schrödinger solitons [4, 20-23] and polarization rotation locked vector Schrödinger solitons [23]. However, accompanying with the development of dispersion management and the goal of ultrashort pulses with high energy, scientific interest of mode-locking operation is gradually shifted into normal-dispersion regime, where spectral filter gets involved in the composite balance [24-27]. These DSs formed in normal-dispersion regime are distinctly characterized by steep spectral edges and large frequency chirp, as well as possessing much higher pulse energy and broader spectral bandwidth than conventional Schrödinger solitons [28-31]. Therefore, developing VDSs in normal-dispersion regime is twofold motivated by both the dynamics unusual from conventional Schrödinger solitons and the improved performance for practical applications.

In particular, polarization rotating and polarization locked VDSs have been demonstrated in a dispersion-managed fiber laser with net normal dispersion [32, 33]. The formation of these two kinds of VDSs is ascribed to a result of coherent coupling between two orthogonally polarized DSs. Moreover, VDSs under incoherent coupling in normal-dispersion regime have been demonstrated as well [34]. The two orthogonally polarized DSs are characterized by large central wavelength disparity and travel together at the same group velocity in the laser cavity, forming the so-called group velocity locked vector dissipative solitons (GVLVDSs). Previous studies are firmly convinced that the formation of VDSs are an intrinsic property of mode-locked fiber lasers in normal-dispersion regime, the features of which are determined by the role of the residual cavity birefringence of fiber lasers [34]. Whereas dispersion management is always introduced to implement the net normal-dispersion mode-locking especially at 1.55 μm, VDSs are invariably obtained in most of normal-dispersion fiber lasers with an extended cavity length, and the repetition rate is limited to only tens of megahertz. Hence, it is desirable to explore whether similar features of VDSs could be maintained in a normal-dispersion short-cavity fiber laser with a high repetition rate, which is significant to both fundamental scientific research and industrial applications of ultra-high-capacity optical communications.

In this paper, we report on an experimental observation of the 129.3 MHz fundamental mode-locked and the 258.6 MHz 2nd-order harmonic mode-locked GVLVDSs in a dispersion-managed short-cavity fiber laser with net normal dispersion. A transmitted semiconductor saturable absorber (SESA) and a wavelength division multiplexer/isolator/tap (WDM/Isolator/Tap) hybrid module are utilized to shorten the cavity length for achieving a high fundamental repetition rate. Assisted with polarization resolved measurement, we can get insight into the vectorial nature of these DSs, including the characteristics of the two orthogonally polarized components, as well as the polarization locked and polarization rotating GVLVDSs respectively observed under fundamental mode-locking and harmonic mode-locking of the fiber laser. To our best knowledge, it is the first time to study the polarization dynamics of high repetition rate DSs in a normal-dispersion mode-locked fiber laser, which could deepen the understanding of the dynamics of DSs.

## 2. Experimental setup

The dispersion-managed fiber laser schematically depicted in Fig. 1 is exploited to investigate the high repetition rate GVLVDSs in net normal-dispersion regime. In the ring cavity configuration, a 0.75 m erbium-doped fiber (EDF, OFS EDF80) with group-velocity dispersion (GVD) of about −32 (ps/nm)/km is used as the gain medium and pumped by a high power 1480 nm Raman fiber laser source. A WDM/Isolator/Tap hybrid module is utilized to shorten the cavity length. A commercial-available transmitted SESA (BATOP, SA-1550-25-2ps-FC/PC) with large absorption of 25%, modulation depth of 15%, non-saturable loss of 10%, saturation fluence of 300 μJ/cm$^2$, damage threshold of 100 MW/cm$^2$, and recovery time of 2 ps is adopted to initialize the mode-locking of the fiber laser. The GaAs SA chip with a thickness of 100 μm is fiber butt-coupled inside a ceramic ferrule, and connected to single-mode fiber patch cables with FC/PC connectors. An in-line fiber polarization controller (PC) is inserted between the EDF and the transmitted SESA to appropriately control the net cavity birefringence. All the pigtails of these components are single-mode fiber (SMF) with GVD of 16 (ps/nm)/km and the total length of them is about 0.85 m. Consequently, the overall length and the net dispersion of the cavity are about 1.60 m and 0.013 ps$^2$ respectively, which are conducive to the formation of GVLVDSs with a high repetition rate. In addition, to have insight into the two orthogonally polarized DSs, a fiber-pigtailed polarized beam splitter (PBS) is connected to the 10% output tap port of the WDM/Isolator/Tap hybrid module. The incoming branch of the PBS is made of SMF, while its two outgoing branches are made of polarization-maintaining fiber (PMF). Moreover, to eliminate the influence of the fiber birefringence introduced by the SMF, a fiber-based PC is inserted between the 10% output tap port and the incoming branch of the PBS to compensate the extra fiber birefringence.

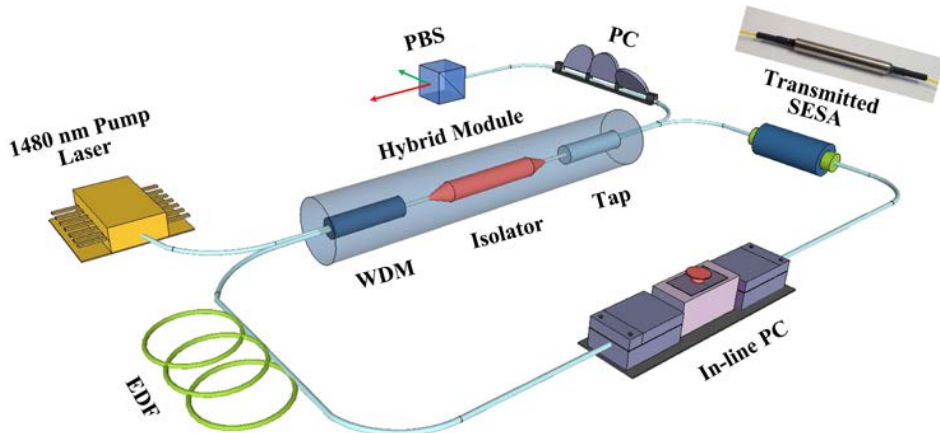

Fig. 1. Schematic of the passively mode-locked fiber laser.

## 3. Experimental results and discussions

In the experiment, self-started mode-locking of the fiber laser can be easily achieved by increasing the pump power above the mode-locking threshold and appropriately setting the PC. The mode-locked pulses are automatically shaped into DSs due to the net normal dispersion of the laser cavity. When the pump power is set at 141 mw, fundamental mode-locking of the fiber laser is obtained as illustrated in Fig. 2. The optical spectrum shown in Fig. 2(a) exhibits a rectangular spectral shape with a

3-dB bandwidth of 7.9 nm, and the autocorrelation trace depicted in Fig. 2(b) declares a pulse width of 3.7 ps with the assumption of a Gaussian pulse shape. Thus, the time-bandwidth product is about 3.7, indicating that the DSs are strongly chirped. The slightly modulated spectrum top is most likely caused by the effective gain which is the combined effect from gain, gain dispersion, component filter etc.. Figures 2(c) and (d) respectively present the oscilloscope trace and radio frequency (RF) spectrum of the DSs. The fundamental repetition rate is fixed at 129.3 MHz according to the RF spectrum, which agrees with the pulse interval of 7.7 ns. The average output power of the fundamental mode-locking is about 5.4 mw, corresponding to pulse energy of 41.8 pJ. The overall lasing efficiency is about 3.8%. The main reason for the low lasing efficiency is principally due to the relatively large insertion loss induced by the SESA with absorption of 25%.

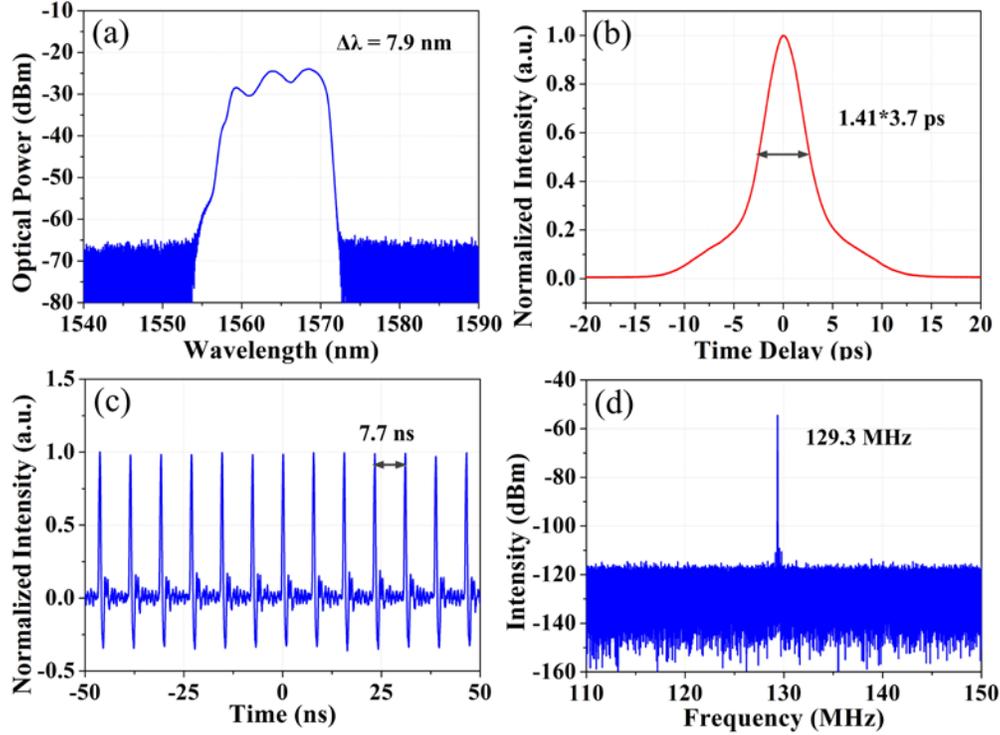

Fig. 2. (a) Optical spectrum, (b) autocorrelation trace, (c) oscilloscope trace and (d) RF spectrum of the fundamental mode-locked DSs.

In consideration of no polarization sensitive components being incorporated in the laser cavity, the formed DSs are intrinsically VDSs, which compose of two orthogonally polarized DSs. To gain insight into the vectorial nature of the fundamental mode-locked DSs formed in normal dispersion regime, we let the output laser pass through the PBS. Figure 3(a) shows the polarization resolved optical spectra of the VDSs. The optical spectra along two orthogonal polarization axes possess distinctly different central wavelengths. From Fig. 2(c), these two orthogonally polarized DSs travel together as a non-dispersive unit in the laser cavity at the same group velocity. Thus, the formed VDSs are the so-called GVLVDSs. Figure 3(b) depicts the oscilloscope traces after the polarization resolved measurement. The pulse intensity of these two orthogonally polarized DSs is mainly uniform, which suggests that the formed GVLVDSs have a fixed state of polarization during propagation in the laser cavity. Moreover, the optical spectrum of the formed GVLVDSs shown in Fig. 3(a) is not characterized by the expected steep spectral edges, especially for the short-wavelength spectral edge, which is different from the optical spectrum of the conventional DSs. Indeed, the optical spectra of the two orthogonally polarized DSs respectively possess steep spectral edges and jointly compose the overall spectrum of the GVLVDSs. However, due to the different central wavelengths and optical power of the two orthogonally polarized components, the superposed optical spectrum has a relatively gradual edge with several small steps instead of the steep edges. For GVLVSs formed in anomalous-dispersion regime with moderate birefringence, two sets of Kelly sidebands usually appear on the optical spectrum, which can be considered as a visualized indicator of GVLVSs. Nevertheless, this visualized indicator is not applicable for GVLVDSs. As a substitution, the aforementioned distinct optical spectrum with gradual spectral edges or several small steps could be considered as an auxiliary indicator of GVLVDSs, which have been also observed in the previous study [34].

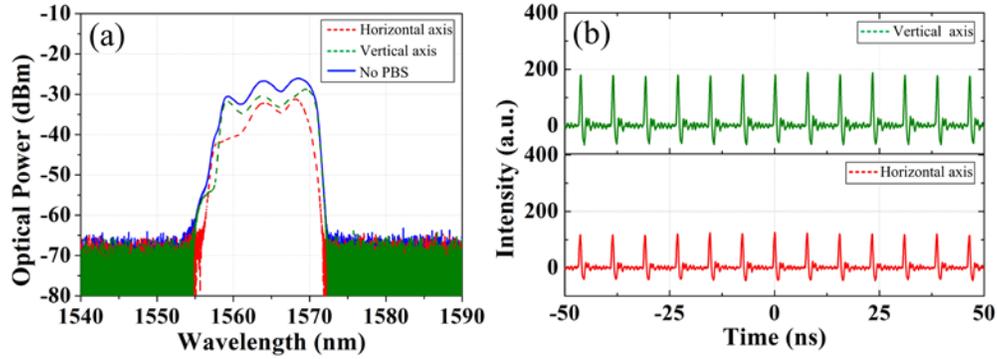

Fig. 3. (a) Polarization resolved optical spectra of the fundamental mode-locked GVLVDSs; (b) oscilloscope traces of these GVLVDSs along the two orthogonal polarizaiton axes.

Furthermore, with increased pump power of 178 mw and appropriate PC setting, 2nd-order harmonic mode-locking with an average output power of 6.3 mw is obtained in the same laser cavity. The optical spectrum shown in Fig. 4(a) is also characterized by a rectangular spectral shape with a 3-dB bandwidth of 7.5 nm. As expected, a small step as the indicator of GVLVDSs appears on the short-wavelength spectral edge. Inset of Fig. 4(a) presents the autocorrelation trace of the GVLVDSs, and the pulse width is estimated to be 3.6 ps with the assumption of a Gaussian pulse shape. The time-bandwidth product can be calculated as 3.4, indicating that the harmonic mode-locked GVLVDSs are chirped as well. Figure 4(b) shows the oscilloscope trace of the GVLVDSs before the PBS, where two pulses exist in one cavity roundtrip time of 7.7 ns with an equal interval, corresponding to a repetition rate of 258.6 MHz. It is demonstrated that GVLVDSs can be still formed with such a high repetition rate. And the pulse energy is reduced to 24.4 pJ on account of the reduplicated repetition rate. Similarly, we also adopt the polarization resolved measurement to get insight into the vectorial nature of the 2nd-order harmonic mode-locked DSs. As shown in Fig. 4(c), the polarization resolved optical spectra along two orthogonal polarization axes also possess different central wavelengths. However, different from the fundamental mode-locked GVLVDSs characterized by a fix polarization, the 2nd-order harmonic mode-locked ones are polarization rotating during propagation in the laser cavity, which can be easily identified by the oscilloscope traces without/after passing through the PBS. Without passing through the PBS, all the pulses have a uniform intensity as shown in Fig. 4(b). On the contrary, from Fig. 4(d) after passing through the PBS, the pulse intensity becomes varying with the roundtrips, straightforwardly manifesting polarization rotating state of the 2nd-order harmonic mode-locked GVLVDSs.

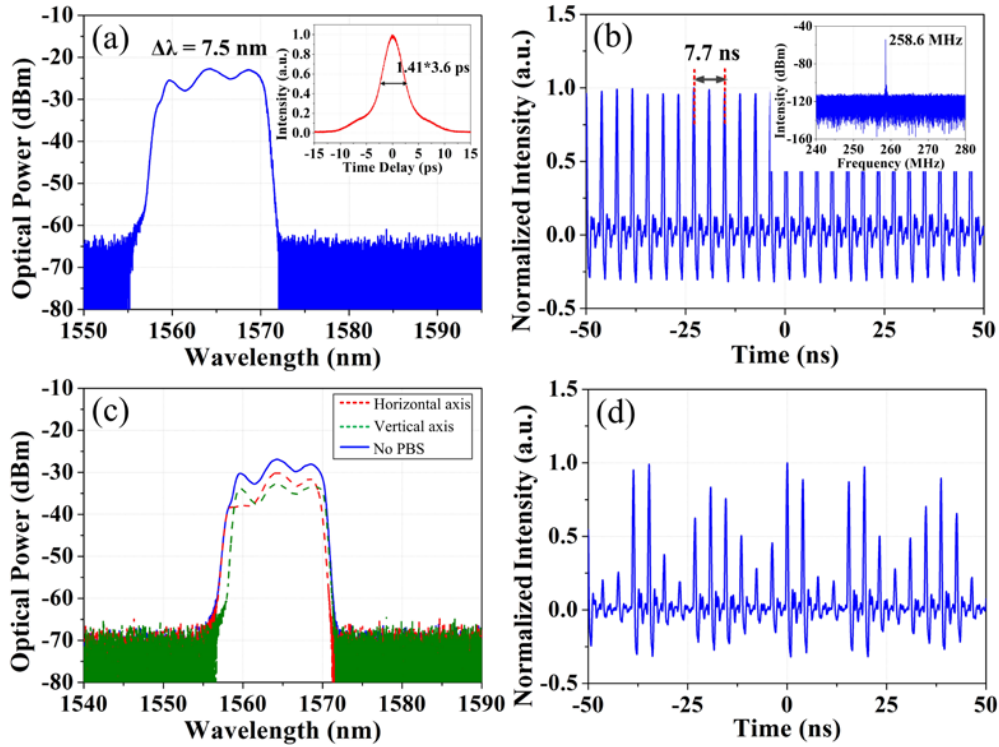

Fig. 4. (a) Optical spectrum and (b) oscilloscope trace of the 2nd-order harmonic mode-locked GVLVDSs (insets of Fig. 4(a) and (b) are respectively the corresponding autocorrelation trace and the RF spectrum); (c) polarization resolved optical spectra of the GVLVDSs; (d) oscilloscope trace of the GVLVDSs after passing through the PBS.

**4. Conclusions**

In conclusion, we report on the first observation of high repetition rate GVLVDSs in a dispersion-managed short-cavity fiber laser with net normal dispersion. The experimental results reveal that, for repetition rate greater than 100 MHz, the two orthogonally polarized DSs with distinctly different central wavelengths can be still coupled together to form the so-called GVLVDSs in normal dispersion regime. In particular, polarization locked and polarization rotating GVLVDSs are respectively observed under the 129.3 MHz fundamental mode-locking and the 258.6 MHz 2nd-order harmonic mode-locking of the fiber laser. Additionally, the distinct optical spectra with a gradual spectral edge and small steps can be a ubiquitous indicator of GVLVDSs. Overall, the investigation on the feasibility and characteristics of the high repetition rate GVLVDSs provides important information for both fundamental scientific research and industrial applications of ultra-high-capacity optical communications.

**Acknowledgments**

This work is supported by the sub-Project of the Major Program of the National Natural Science Foundation of China (N0.61290315), the National Key Scientific Instrument and Equipment Development Project of China (No. 2013YQ16048707), and the National Natural Science Foundation of China (No. 61275004, No. 61275109).

**References and links**

1. P. Grelu and N. Akhmediev, "Dissipative solitons for mode-locked lasers," Nat. Photonics 6, 84–92 (2012).
2. N. Akhmediev, and A. Ankiewicz (Eds.), "Dissipative Solitons: From optics to biology and medicine," Lecture Notes in Physics, Springer, Berlin-Heidelberg (2008).
3. B. Oktem, C. Ülgüdür, F. Ö. Ilday, "Soliton–similariton fibre laser," Nat. Photonics 4, 307–311 (2010).
4. S. Cundiff, B. Collings, and W. Knox, "Polarization locking in an isotropic, modelocked soliton Er/Yb fiber laser," Opt. Express 1, 12-21 (1997).
5. J. W. Haus, G. Shaulov, E. A. Kuzin, and J. Sanchez-Mondragon, "Vector soliton fiber lasers," Opt. Lett. 24, 376–378 (1999).
6. B. Collings, S. Cundiff, N. Akhmediev, J. Soto-Crespo, and K. Bergman, et al, "Polarization-locked temporal vector solitons in a fiber laser: experiment," JOSA B 17, 354–365 (2000).

7. D. Y. Tang, H. Zhang, L. M. Zhao, and X. Wu, "Observation of High-Order Polarization-Locked Vector Solitons in a Fiber Laser," Phys. Rev. Lett. 101, 153904 1–4 (2008).
8. K. Tamura, E. P. Ippen, H. A. Haus, and L. E. Nelson, "77-fs pulse generation from a stretched-pulse mode-locked all-fiber ring laser," Opt. Lett. 18, 1080-1082 (1993).
9. T. F. Carruthers, I. N. Duling, M. Horowitz, and C. R. Menyuk, "Dispersion management in a harmonically mode-locked fiber soliton laser," Opt. Lett. 25, 153-155 (2000).
10. J. W. Nicholson and M. Andrejco, "A polarization maintaining, dispersion managed, femtosecond figure-eight fiber laser," Opt. Express 14, 8160-8167 (2006).
11. N. N. Akhmediev, A. Ankiewicz, and J. M. Soto-Crespo, "Stable soliton pairs in optical transmission lines and fiber lasers," JOSA B 15, 515–523 (1998).
12. A. Zavyalov, R. Iliew, O. Egorov, and F. Lederer, "Dissipative soliton molecules with independently evolving or flipping phases in mode-locked fiber lasers," Phys. Rev. A 80, 043829 1–8 (2009).
13. B. Ortac, A. Zavyalov, C. K. Nielsen, O. Egorov, and R. Iliew, et al, "Observation of soliton molecules with independently evolving phase in amode-locked fiber laser," Opt. Lett. 35, 1578–1580 (2010).
14. Y. Y. Luo, Q. Z. Sun, L. M. Zhao, Z. C. Wu, Z. L. Xu, S. N. Fu, and D. M. Liu, "Bound states of group-velocity locked vector solitons in a passively mode-locked fiber laser," in Asia Communications and Photonics Conference 2015, (Optical Society of America, 2015), ASu1C.3.
15. W. Chang, A. Ankiewicz, J. M. Soto-Crespo, and N. Akhmediev, "Dissipative soliton resonances," Phys. Rev. A 78, 023830 (2009).
16. L. Duan, X. M. Liu, D. Mao, L. R. Wang, and G. X. Wang, "Experimental observation of dissipative soliton resonance in an anomalous-dispersion fiber laser," Opt. Express 20, 265-270 (2012).
17. P. Serena, N. Rossi, and A. Bononi, "PDM-iRZ-QPSK vs. PS-QPSK at 100 Gbit/s over dispersion-managed links," Opt. Express 20, 7895–7900 (2012).
18. L. M. Zhao, D. Y. Tang, H. Zhang, X. Wu, and N. Xiang, "Soliton trapping in fiber lasers," Opt. Express 16, 9528-9533 (2008).
19. L. Yun, X. M. Liu, and D. D. Han, "Observation of vector-and scalar-pulse in a nanotube-mode-locked fiber laser," Opt. Express 22, 5442-5447 (2014).
20. S.V. Sergeyev, C. B. Mou, A. Rozhin, and S. K. Turitsyn, "Vector Solitons with Locked and Precessing States of Polarization," Opt. Express 20, 27434–27440 (2012).
21. C. B. Mou, S. V. Sergeyev, A. G. Rozhin, and S. K. Turitsyn, "All-fiber polarization locked vector soliton laser using carbon nanotubes," Opt. Lett. 36, 3831–3833 (2011).
22. C. B. Mou, S. V. Sergeyev, A. G. Rozhin, and S. K. Turitsyn, "Bound state vector solitons with locked and precessing states of polarization," Opt. Express 21, 26868-26875 (2013).
23. V. Tsatourian, S. V. Sergeyev, C. B. Mou, A. Rozhin, V. Mikhailov, B. Rabin, P. S. Westtbrook, and S. K. Turitsyn, "Polarisation dynamics of vector soliton molecules in mode locked fiber laser," Sci. Rep. 3, 3154 (2013).
24. L. M. Zhao, D. Y. Tang, and J. Wu, "Gain-guided soliton in positive group dispersion fiber lasers," Opt. Lett. 31, 1788-1790 (2006).
25. L. M. Zhao, D. Y. Tang, T. H. Cheng, and C. Lu, "Gain-guided solitons in dispersion-managed fiber lasers with large net cavity dispersion," Opt. Lett. 31, 2957-2959 (2006).
26. A. Chong, J. Buckley, W. Renninger, and F. Wise, "All-normal-dispersion femtosecond fiber laser," Opt. Express 14, 10095-10100 (2006).
27. F. W. Wise, A. Chong and W. H. Renninger, "High-energy femtosecond fiber lasers based on pulse propagation at normal dispersion," Laser Photonics Rev. 2, 58-73 (2008).
28. V. L. Kalashnikov, and A. Chernykh, "Spectral anomalies and stability of chirped-pulse oscillator," Phys. Rev. A 75, 033820 (2007).
29. A. Chong, W. H. Renninger, and F. W. Wise, "All-normal-dispersion femtosecond fiber laser with pulse energy above 20 nJ," Opt. Lett. 32, 2408-2410 (2007).
30. K. Kieu, W. H. Renninger, A. Chong, and F. W. Wise, "Sub- 100 fs pulses at watt-level powers from a dissipative-soliton fiber laser," Opt. Lett. 34, 593-595 (2009).
31. X. M. Liu, "Numerical and experimental investigation of dissipative solitons in passively mode-locked fiber laser with large net-normal-dispersion and high nonlinearity," Opt. Express 25, 22401-22416 (2009).
32. H. Zhang, D. Y. Tang, L. M. Zhao, X. Wu, and H. Y. Tam, "Dissipative vector solitons in a dispersion managed cavity fiber laser with net positive cavity dispersion," Opt. Express 16, 455-460 (2009).
33. H. Zhang, D. Y. Tang, L. M. Zhao, Q. L. Bao, and K. P. Loh, "Vector dissipative solitons in graphene mode locked fiber lasers," Opt. Commun. 283, 3334-3338 (2010).
34. L. M. Zhao, D. Y. Tang, X. Wu, and H. Zhang, "Dissipative soliton trapping in normal dispersion fiber laser," Opt. Lett. 35, 1902-1904 (2010).